%% file: main.tex
\documentclass[runningheads]{llncs}

\usepackage{hyperref}
\usepackage{graphicx}
\usepackage[utf8]{inputenc}
\usepackage{amsmath}
\usepackage{threeparttable}
\usepackage{array}
\usepackage{multirow}
\usepackage{xcolor}
\usepackage{color, colortbl}
\newcolumntype{P}[1]{>{\centering\arraybackslash}p{#1}}

\definecolor{Gray}{gray}{0.9}
\definecolor{LightCyan}{rgb}{0.88,1,1}

\begin{document}
\title{Optimizing CNN-based Hyperspectral Image Classification on FPGAs}

\newcommand*\samethanks[1][\value{footnote}]{\footnotemark[#1]}
\author{
Shuanglong Liu\inst{1}\thanks{The first two authors contributed equally.}  \and Ringo S.W. Chu\inst{2}\samethanks \and Xiwei Wang\inst{3} \and Wayne Luk\inst{1}
}
\authorrunning{S.Liu et al.}
\institute{
Department of Computing, Imperial College London, London, UK\\
\email{\{s.liu13, w.luk\}@imperial.ac.uk}\and 
Department of Computer Science, University College London, London, UK \\
\email{ringo.chu.16@ucl.ac.uk} \and 
China Academy of Space Technology, Beijing, China\\
\email{wangxiwei@gmail.com}
}

\maketitle
\input{Abstract.tex}
\input{Introduction.tex}

\input{Background.tex}

\input{Models.tex}

\input{Implementation.tex}

\input{Evaluation.tex}

\input{Summary.tex}
\bibliographystyle{splncs04}
\bibliography{reference}
\end{document}

%% file: Abstract.tex
\begin{abstract}

Hyperspectral image (HSI) classification has been widely adopted in applications involving remote sensing imagery analysis which require high classification accuracy and real-time processing speed.  Methods based on Convolutional neural networks (CNNs) have been proven to achieve state-of-the-art accuracy in classifying  HSIs. However, CNN models are often too computationally intensive to achieve real-time response due to the high dimensional nature of HSI, compared to traditional methods such as Support Vector Machines (SVMs). Besides, previous CNN models used in HSI are not specially designed for efficient implementation on embedded devices such as FPGAs. This paper proposes a novel CNN-based algorithm for HSI classification which takes into account hardware efficiency. A customized architecture which enables the proposed algorithm to be mapped effectively onto FPGA resources is then proposed to support real-time  on-board classification with low power consumption.
Implementation results show that our proposed accelerator on a Xilinx
Zynq 706 FPGA board achieves more than 70$\times$ faster than an Intel 8-core Xeon CPU  and 3$\times$ faster than an NVIDIA GeForce 1080 GPU. Compared to previous SVM-based FPGA accelerators, we achieve comparable processing speed but provide a much higher classification accuracy.

\keywords{Hyperspectral Image Classification \and Deep Learning \and Convolution Neural Network \and Field-Programmable Gate Array}

\end{abstract}

%% file: Introduction.tex
\section{Introduction}\label{sec:intro}

Hyperspectral images (HSI) contain spectrum information for each pixel in the image of a scene, and can be used in finding objects and identifying materials or detecting processes \cite{grahn2007techniques}. Hyperspectral images are widely employed in many applications from airborne and satellite remote sensing mission \cite{arc_ws}, to oil spill detection \cite{salem2001hyperspectral}, early cancer diagnosis \cite{HSICancer} and environmental monitoring \cite{HSIEnviroMonitor}. 
HSI classification involves assigning a categorical class label to each pixel in the image, according to the corresponding spectral and/or spatial feature \cite{bioucas2013hyperspectral}.
With the advent of new hyperspectral remote sensing instruments and their increased temporal resolutions, the availability and dimensionality of hyperspectral data are continuously increasing \cite{lopez2013promise}. This demands very fast processing solutions for on-board space platforms in order to reduce download bandwidth and storage requirements \cite{arc_ws}, making reconfigurable hardware such as FPGAs very promising to perform and accelerate HSI classification methods.


Among the approaches explored for HSI classification, convolutional neural network (CNN) based methods such as BASS Net \cite{santara2017bass} and HSI-CNN \cite{luo2018hsi} are favourable over
the others because of their greatly improved accuracy for some popular benchmark datasets, with the ability to use extensive parameters to learn
spectral features of a HSI. However, these CNN-based algorithms have great computational complexity due to the large dimensionality of hyperspectral images. Besides, prior CNN models used in HSI classification may not be hardware efficient to be deployed on embedded systems such as FPGAs without any modifications since they are not specially designed for FPGAs.

In order to address the above challenges and achieve fast processing speed on embedded devices, this work proposes a novel CNN architecture based on BASS Net \cite{santara2017bass}, and our model is more hardware efficient for implementation on FPGAs while maintaining similar accuracy as the original BASS Net. Besides, we propose and optimize the hardware architecture to accelerate our proposed network in FPGA by parallel processing, data pre-fetching and design space exploration. Compared to previous SVM-based FPGA accelerators, the proposed accelerator has almost the same scale of processing speed on the same scale of FPGA device, but provides a lot higher accuracy results.

The main contributions of this work are summarized as follows:
\begin{itemize}

  \item A novel network for HSI classification which takes into account hardware efficiency, and thus achieves real-time on-board HSI classification with both high accuracy and fast processing speed (Section \ref{sec:net});
  
  \item A highly optimized hardware architecture which maps the proposed CNN model onto FPGAs, and it processes all the layers in on-chip memories to enable high throughput of real-time HSI applications (Section \ref{sec:imp});
  
  
  \item Evaluation of the proposed accelerators on a Xilinx ZC706 FPGA board across four popular benchmark datasets. Our accelerator achieves an overall classification accuracy of 95.8\%, 99.4\%, 95.2\% and 98.2\% respectively which largely outperforms previous SVM-based FPGA accelerators, and it achieves around 10 to 25 us/pixel processing speed which is about 80$\times$ and 3$\times$ faster than the respective CPU and GPU designs (Section \ref{sec:eva}).
\end{itemize}

%% file: Background.tex
\section{Background and Related Work}
\subsection{Hyperspectral Imagery}



Unlike traditional RGB image, hyperspectral images are typically represented as a data cube in dimension $(x, y, \lambda)$, where $x$ and $y$ represent spatial dimensions with space information of pixels, and $\lambda$ represents the third dimension with spectral information for distinguishing different materials and objects. 

Hyperspectral image (HSI) classification is the task to assign a class label
to every pixel in an image. Several approaches have been explored in literature for HSI classification. K-nearest neighbors (k-NN) based methods use Eucledian distance in the input space to find the k nearest training examples and a class is assigned on the basis of them \cite{santara2017bass}. Support Vector Machine (SVM) based methods introduce dimensionality reduction in order to address the problem of high spectral dimensionality and limited number of labeled training examples, with SVM classifiers used in the reduced dimensional space. Although these methods adopt parallel processing \cite{arc_ws} and are suitable for FPGA-based acceleration, they often behave weakly in terms of the classification accuracy when tackling large datasets \cite{santara2017bass}.


\subsection{CNN-Based HSI classification}

Recently, deep learning based methods have achieved promising performance in HSI classifications \cite{chen2014deep} due to their ability to use extensive parameters to learn features. Deep learning methods \cite{zhang2016deep} utilize spectral-spatial context modeling in order to address the problem of spatial variability of spectral signatures. These methods often use convolutional neural networks (CNNs) for feature learning and classification in an end-to-end fashion. CNNs adopt extensive parameter-sharing to tackle the curse of dimensionality. They extract and learn representative features via multiple times of back propagation. Using features is more effective than rule-based algorithms for recognition tasks.

One of the most popular CNN models for HSI classification: BASS Net \cite{santara2017bass}, a deep neural network architecture
which learns band-specific spectral-spatial features and gives
state-of-the-art performance without any kind of data-set
augmentation or input pre-processing. While this algorithm leads to high classification performance due to efficient band-specific feature learning, the model is computationally intensive, which often requires huge amount of resources and energy. Nevertheless, this network has parallelism in many computational blocks and thus can be prallelized in hardware platforms such as FPGAs. 
However, the BASS Net is not suitable to be deployed on embedded systems without modification. The main challenge is that the CNN architecture does not have identical layer parameters, which increases the difficulty of designing generic hardware modules that support varying parameters.
For example, there are 1-D convolutional layers with different kernel sizes such as 3$\times$1 and 5$\times$1 implemented using spectral information and 2-D convolutional layers applied using spatial information. Fully-connected layers are also applied after all convolution layers for summarization and output classification probability.
Because of these reasons, direct mapping of this algorithm to FPGAs may not be efficient and cannot satisfy the requirement of real-time processing without the proposal of algorithm adaptions and efficient hardware architecture.

\subsection{Related Work}



Prior work includes deploying SVM-based HSI classification on FPGA for acceleration and utilizing GPUs for algorithmic speed up on CNN-based methods. There is exhaustive literature on accelerating the traditional algorithms such as SVMs using FPGAs. Wang et al. \cite{arc_ws} proposed a novel accelerator architecture for real-time SVM classification. The accelerator uses data flow programming to
achieve high performance and can be used for different applications.
Tajiri et al. \cite{fpt2018} proposed a hyperspectral image
classification system on FPGA, by introducing the Composite
Kernel Support Vector Machine and reducing the computational complexity.
These former accelerators achieve real time processing speed but they do not achieve high classification accuracy and therefore are not favoured over CNN-based methods.

Recently CNN-based HSI approaches have been proposed by many researchers. Santara et al. \cite{santara2017bass} presented an end-to-end deep learning
architecture that extracts band specific spectral-spatial features
to performs landcover classification. 
Luo et al. \cite{luo2018hsi} proposed a novel HSI classification model to reorganize data by using the correlation between convolution results and to splice the one-dimensional data into image-like two-dimensional data to deepen the network structure and enable the network to extract and distinguish the features better.
Lee et al. \cite{lee2017going} built a fully convolutional
neural network with a total of 9 layers, which is much deeper than other convolutional networks for HSI classification. To enhance the learning efficiency of the proposed network trained on relatively sparse training samples, residual learning was used in their work.
However, all of these efforts have only focused on the improvement of the accuracy of these algorithms on CPU or GPUs, the performance of their models have never been reported or considered in prior works. Therefore, it is unclear if these algorithms are suitable for on-board platforms and it is  not straightforward to map them into embedded devices for real-time processing.

To the best of the authors' knowledge, this is the first work that proposes FPGA architecture for CNN-based HSI classifications. Our FPGA-based accelerator achieves high accuracy, fast processing speed and lower power consumption, which is suitable for on-board space platforms. 

%% file: Models.tex
\section{Proposed CNN-based HSI Classification Model}\label{sec:net}

\subsection{HSI Preprocessing method}
Deep neural network architectures are based on pixel-wise classification results of the input image. More explicitly, the input of the network is a $p\times p\times N_{c}$ pixel cube using its central pixel $x$ as label, where $p$ refers to the and $N_{c}$ is the number spectral channels of the dataset scene. The output of the network is the predicted class label $y_{i}$ for pixel $x$.

\begin{figure}[ht]
\includegraphics[scale=0.12]{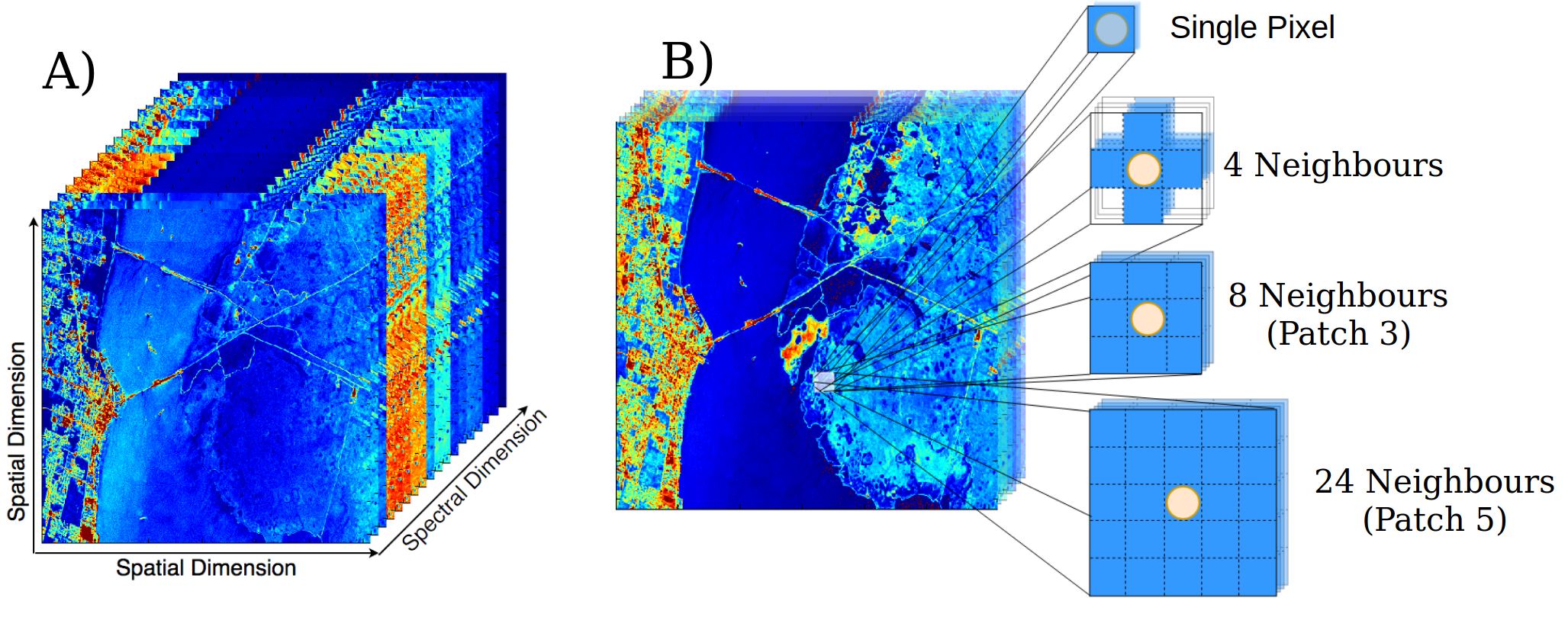}
\centering
\caption{A) Hyperspectral Image cube. B) Extraction of data cube with labeled at central pixel from a raw HSI.}
\label{fig:HSIandpreprocess}
\end{figure}

The patched based method was first introduced by Leng et al \cite{Leng2016CubeCNNSVMAN} to analysis neighbour spatial information around a pixel. Different strategies are described in Figure \ref{fig:HSIandpreprocess}.

\subsection{Model Structure Description}

We propose our network based on the structure of BASS-Net \cite{santara2017bass}. The network uses significantly fewer parameters when compared to other models and exhibits parallelism at inference stage. We further modify the model such that it is more efficient for FPGA implementation. The patch strategy is altered to 24 neighbours (Patch 5) (see Figure \ref{fig:HSIandpreprocess}) and replacing 1-D kernels with 2-D kernels for the purpose of achieving better HSI classification performance.

\begin{figure}[ht]
\includegraphics[scale=0.185]{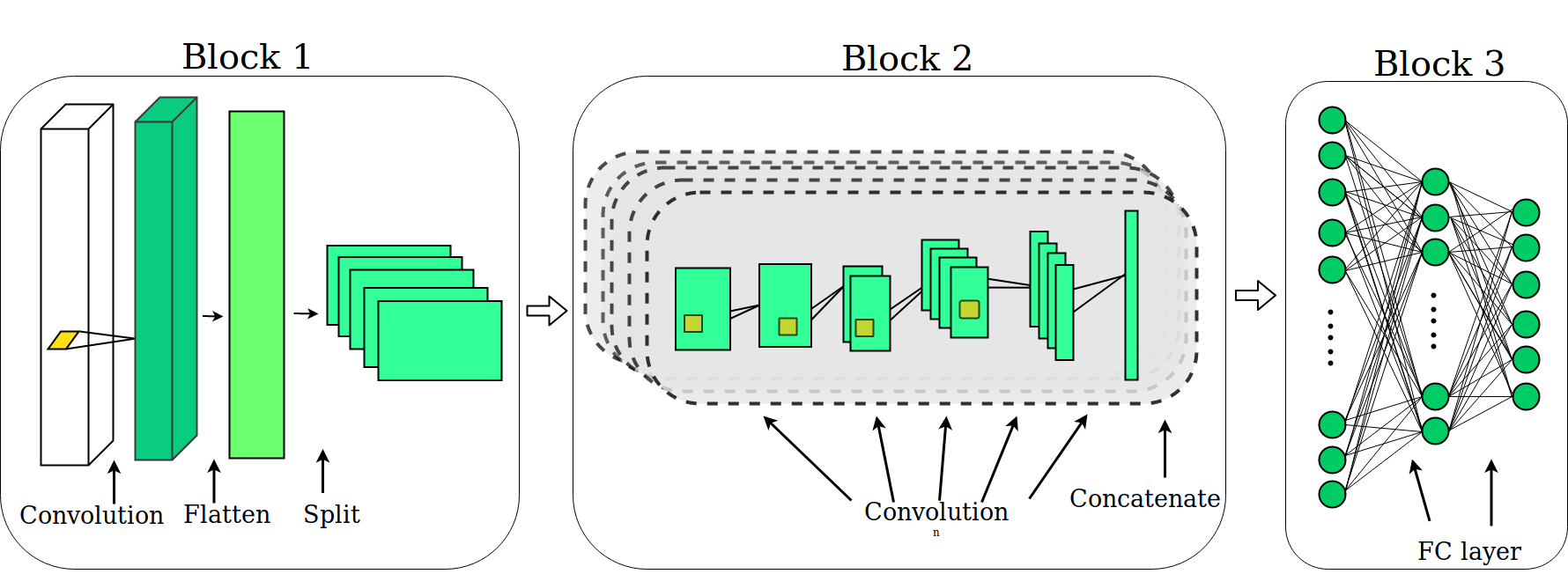}
\centering
\caption{Overall structure of our model architecture.}
\label{fig:PSSRNET}
\end{figure}

The proposed model processes through the following three stages as shown in Figure \ref{fig:PSSRNET}.

\paragraph{Spectral feature selection and band partitioning (Block 1)}
In this step, the $p \times p \times N_{c}$ input volume is taken as input by a 3 $\times$ 3 or 1 $\times$ 1 spatial convolution for feature selection; then the spectral dimension of the output is split into $N_{b}$ bands with equal bandwidth and passed as input to the second step for parallel processing;

\paragraph{Spectral feature learning (Block 2)}
 
This step applies $N_b$ parallel networks, with one for each band: the input is  first flattened  to  one  dimension  along  the  spatial  dimensions; four $3 \times 3 $  convolution filter is applied in the spectral dimension to learn spectral features; the outputs of the parallel networks are then concatenated and fed into the summarization and classification stage. Rectified Linear Unit (ReLu) is applied on every layers to accelerate convergence. 


\paragraph{Summarization and Classification (Block 3)}
This step summarizes the concatenated output of the band-specific networks of the previous stage by using a set of fully connected layers, each of which is followed by a ReLu layer A K-way Softmax layer performs the final classification by calculating the conditional probabilities of the K output classes.


\begin{table}[!ht]
    \centering
     \caption{The proposed CNN-based HSI network with input patch size $5 \times 5 $ and $N_{b}=4$.}
    \begin{tabular}{|c|c|c|c|c|c|}
    \hline
    \multicolumn{6}{ c } {Input patch size: 5 $\times$ 5 $\times$ 220 and $N_{b}=4$} \\ \hline
         \multicolumn{2}{ c| } {Layer} & type  & Input Volume & Output Volume & kernel  \\ \hline
         Block 1 & layer 1 & 3 $\times$ 3 conv & 5 $\times$ 5 $\times$ 220 & 3$\times$3$\times$220 & 3$\times$3$\times$220$\times$220 \\ \hline
        
         \multicolumn{6}{ c } {split into $N_{b}$ bands along spectral dimension and each is run in Blcok 2} \\ \hline 
         
         \multirow{4}{*}{ Block 2} & layer 1 & 3 $\times$ 3 conv & 9$\times$55$\times$1 & 7$\times$53$\times$2 & 3$\times$3$\times$1$\times$2 \\ 
        & layer 2 & 3 $\times$ 3 conv & 7$\times$53$\times$2 & 5$\times$51$\times$4 & 3$\times$3$\times$2$\times$4 \\ 
        & layer 3 & 3 $\times$ 3 conv & 5$\times$51$\times$4 & 3$\times$49$\times$4 & 3$\times$3$\times$4$\times$4 \\ 
        & layer 4 & 3 $\times$ 3 conv & 3$\times$49$\times$4 & 1$\times$47$\times$4 & 3$\times$3$\times$4$\times$4 \\ \hline
        
        \multicolumn{6}{ c } {The output of $N_{b}$ Blocks concatenated} \\ \hline 
         
         \multirow{2}{*}{Block 3} & layer 1 & Fully Connect layer & 752$\times$1 & 120$\times$1 & 752$\times$120 \\
       & layer 2 & Fully Connect layer & 120$\times$1 & 9$\times$1 & 120$\times$9 \\ \hline
    \end{tabular}
    \label{tab:net}
   
\end{table}

\subsection{Hardware Adoptions}

There are some tunable parameters in the network architecture for different design choices:the input patch size $p$, the number of parallel bands $N_{b}$ in the second stage and the convolutional kernel in the first stage. Compared to BASS Net, three major changes are introduced for hardware efficiency and accuracy improvement: 
\begin{itemize}
    \item[1)] We utilized 3 $\times$ 3 CNN filters in  stage 1 for spatial dimension learning, which amplifies spatial signatures and are identical to the convolutions in stage 2;
    \item[2)] 1-D convolutions of kernel size 3$\times$1 and 5$\times$1 in stage 2 are all replaced by 2-D 3$\times$3 fixed kernel size for generic hardware module design and reuse;
    \item[3)] In stage 2, the data are flattened along the spatial dimension and split into $N_{b}$ segments along the spectral dimension.  We choose $N_{b}$ from 2, 4 or 8 for easy parallel processing in FPGA technology. 
\end{itemize}
For block 1, we also apply 3$\times$3 convolution for input patch size of 5$\times$5$\times N_{c}$ or 1$\times$1 convolution for input patch size of 3$\times$3$\times N_{c}$. One of the strategies is chosen for different datasets in order to have a trade-off between the accuracy and processing speed.

Our final network\footnote{https://github.com/custom-computing-ic/CNN-Based-Hyperspectral-Image-Classification} parameter configurations are summarized in Table \ref{tab:net}.

\subsection{Training Process}

\paragraph{Regularization methods}
Trainable parameters are shared across each band in block 2 in order to minimize hardware resources and design space. Bands must be placed sequentially to allow back-propagation and avoid gradient vanishing problem. Dropout is applied on the fully-connected layers in block 3 to prevent over-fitting. 

\paragraph{Loss functions}
We employ cross-entropy loss function as error measure. The training process aims to minimise the loss value to obtain a distribution that is the best capture of the data set. Given a training dataset: $\{X_i, y_i\}_{i=1}^N$, the designated loss function is described as:

    \begin{equation} \label{eq:lossfun}
\mathcal{L}(p, \hat{p}) = -\sum_{i=1}^{N}p(x)\log\Hat{p}(x) 
\end{equation}

where $p(x)$ is the probability distribution of our models, and $\Hat{p}(x)$ is the actual distribution that represents the dataset.

\paragraph{Optimizer}
We used Adaptive Moment Estimation to update kernel weights and biases with initial learning rate 0.0005, $\beta_1$ = 0.9 and  $\beta_2$ = 0.999

%% file: Implementation.tex
\section{Proposed CNN-based HSI Accelerator}\label{sec:imp}

\subsection{Hardware Architecture}

\begin{figure}[ht]
     \centering
     \includegraphics[width=0.8\textwidth]{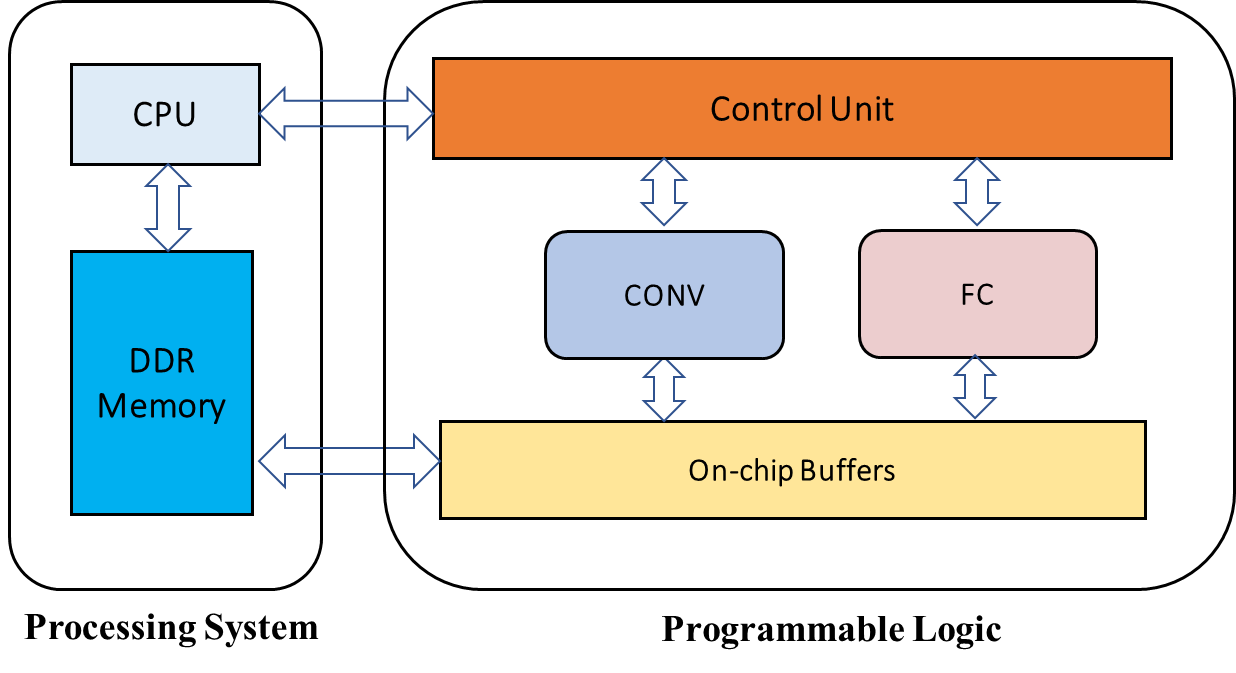}  
      
    \caption{The architecture of the FPGA-based CNN accelerator which integrates 2 computation units: CONV and FC.}
  
\label{fig:hw_top}
\end{figure}

Based on our proposed network, we design the hardware architecture of the FPGA accelerator for CNN-based real-time HSI classification shown in Figure \ref{fig:hw_top}. The proposed CNN accelerator design on FPGA is composed of several major components: the computation units (CONV and FC modules), on-chip buffers, external memory and
control unit. The processor (PS) configures the parameters of the layers for the two computation units through the control unit. 
CONV and FC are the basic computation units for the CNN-based HSI classification algorithm. Due to the limitation of the on-chip memory size, the input data and weights of all the layers are stored in off-chip memories and transferred to on-chip buffers when processing each layer.
All intermediate data for processing are stored in
on-chip buffers to avoid frequent off-chip memory access. Therefore, the required on-chip buffers need to store at least the input and output data size of one layer, since these on-chip buffers will be reused when the next layer is processed.

\subsection{Convolution and Fully-Connected Design}

Our convolutional and fully-connected layer architectures are inspired by the design in \cite{zhao2017optimizing}. The CONV unit contains several computational kernels running
in parallel, and each kernel consists of 9 multipliers followed by an adder tree to implement the 3$\times$3 2-D convolution operation.
Multiple kernels are utilised for parallel channel and filter processing and the total number of kernels represents the parallelism of the CONV unit ($P_{C}$). Besides, we implement 1$\times$1 convolutions in the CONV kernel by reusing the 9 multipliers and bypassing the adder trees. Therefore, the degree of parallelism of 1$\times$1 CONV is 9 times of $P_{C}$.

The operation of the FC kernel is to perform dot product between the reshaped
input feature vector and the weight matrix. The FC kernel also contains several multipliers to calculate the dot product between each row of the weight
matrix and the feature vector in parallel. The number of multipliers in the FC kernel represents the parallelism of the FC unit ($P_{F}$).

\subsection{Optimizations}

In this section, we describe the optimization techniques used for the FPGA-based HSI accelerator design in order to increase the system throughput. The optimizations to the FPGA-based accelerator mainly focus on: 1) fully utilising the existing hardware
resource to reduce the computation time by parallel processing \cite{liu2014parallel}, and 2) increasing the data reuse to reduce the communication time to off-chip memories\cite{liu2017communication}. 

\subsubsection*{Data pre-fetching and pipelining.}

\begin{figure}[ht]
     \centering
     \includegraphics[width=0.9\textwidth]{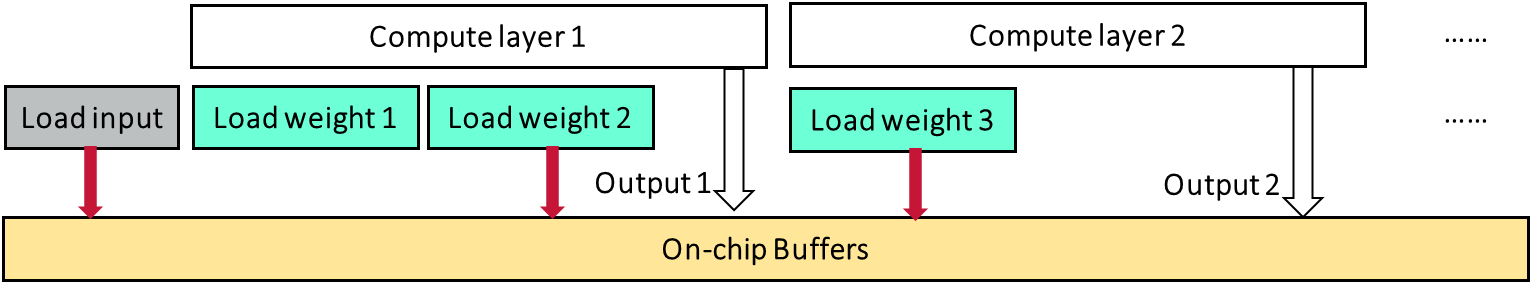}  
     
    \caption{Computation and weights transfer flows.}
 
\label{fig:prefetch}
\end{figure}

We adopt the parallel processing of multiple data and filters inside the CONV and FC kernel to increase the data reuse and reduce computation overhead. However, there is another overhead involving the transfer of the weights from DDR memory to the computational units. These weights actually do not need to be stored in on-chip buffers as they are only used once which is different from input data. To reduce this overhead, weights are pre-fetched before processing in order to overlap weight transfer time and computation. Figure \ref{fig:prefetch} shows the timing of several computation and weight transfer phases.

To compute the first layer, we first load input data and weights of layer 1 and 2 from DDR memory to on-chip buffers. At the same time, the computation of layer 1 can start as soon as the weights of layer 1 are valid since the input data are already in on-chip buffers; after the computation of layer 1, the output of layer 1 has already been stored in the on-chip buffers and the weights of layer 2 have loaded to the memories, so we can process layer 2 immediately  after finishing layer 1 and at the same time we load the weights for the next stage, i.e., layer 3. As a result, the total execution time only needs to cover the transfer of the input and the final output, and the computational time. All the weight transfer time is overlapped in the computation stage, and there is no waiting time between computations of two consecutive layers.

\subsubsection*{Data Quantization.}

The main benefit of accelerating CNN models in 
FPGAs comes from the fact that CNNs are robust to low bitwidth quantization \cite{liu2017trets}. Instead of
using the default double or single floating point precision in
CPU, fixed-point precision can be used in FPGA-based CNN
accelerator to achieve an efficient design optimized for performance
and power efficiency \cite{liu2015exact,liu2017unbiased}. In this work, we implement our proposed design with 16 bit fixed-point which has been shown to achieve almost the same accuracy as floating point in the inference stage, in order to allow optimizations for high
parallelism mentioned in the above section. 
It should be noted that there is no significant accuracy loss in the HSI classification result when reducing the precision from  32-bit floating point to  16-bit fixed-point quantized version for the inference process, as long as the training stage adopts 32-bit floating point.


\subsubsection*{Design Parameter Tuning.}

We tune the hardware design parameters of the accelerator mentioned above based on the network input size and structure, in order to fully utilize the computation resources (DSPs) and achieve the optimal performance for the proposed CNN accelerator. This process involves adjusting the computational resource allocations, i.e., $P_{C}$ and $P_{F}$ between the CONV unit and FC unit to achieve minimal computation time.

This step in essence covers design space exploration (DSE).
The design parameters used in the network and the hardware accelerator are summarized in Table \ref{table:parameters}. In our approach, we first adjust the network parameters in the training stage to verify the accuracy results. Then for a given set of network parameters, we develop a tool with the Nonlinear programming solver \textit{fmincon} in Matlab Optimization Toolbox to automatically generate the optimal hardware design parameters in terms of the processing speed. 
Based on our approach, we can easily extend our network to support different HSI datasets and achieve the corresponding optimal speed. Therefore, it largely improves the design quality and designer productivity.

\begin{table}[ht]
\centering
 
\caption{Network and hardware accelerator parameters for design space exploration.}
\label{table:parameters}

\begin{tabular}{| P{3cm} | p{7cm} |} 

\hline

\multicolumn{2}{ |c| }{Data-set variables}  \\ \hline
$C$ & number of classes in the HSI dataset \\ \hline
$N_{c}$ & spectral bands or input volume channels \\ \hline

\multicolumn{2}{ |c| }{Tunable network design parameters}  \\ \hline
$N_{b}$ &  number of split band in Block 2  \\ \hline
$ps$  & input patch size    \\ \hline

\multicolumn{2}{ |c| }{Tuable hardware design parameters}  \\ \hline
$P_{C}$ &  Parallelism of CONV unit  \\ \hline
$P_{F}$  & Parallelism of FC unit    \\ \hline

\end{tabular}

\end{table}

%% file: Evaluation.tex
\section{Evaluation}\label{sec:eva}

In this section, the accuracy of our proposed network is compared to other CNN-based algorithms and some traditional methods. The performance of our accelerator is also compared to prior FPGA-based accelerators.

\subsection{Benchmarks}
\label{subsec:bench}

Four benchmark datasets\footnote{These datasets can be obtained from  \url{http://www.ehu.eus/ccwintco/index.php?title=Hyperspectral_Remote_Sensing_Scenes}.} are used to evaluate our proposed model and accelerator. These include Indian Pines scene, Salinas scene, Kennedy Space Centre  (KSC) scene and Botswana scene. The first three datasets were acquired by the NASA Airborne Visible Infra-Red Imaging Spectrometer (AVIRIS) across Indiana, California, Florida, and the Botswana dataset was acquired by the NASA EO-1 satellite over Botswana. The spectral bands used for these datasets after removing bands covering the region of water absorption are 220, 224, 176 and 144 respectively. 
The classes and spectral bands of each dataset are summarized in Table \ref{tab:config}.
Some classes are dropped during training due to limited samples. We randomly select 15\% as training samples, 5\% as validation samples and the reminder as testing samples. 
The network parameter configurations after tunning in the training stage are  also summarized in Table \ref{tab:config}.
\begin{table}[h]
    \centering
     \caption{Dataset variables and their corresponding network configurations.}
    \begin{tabular}{|P{2cm}|P{1cm}|P{2.5cm}|P{1.5cm}|P{1cm}|P{2cm}|}
    \hline
    Dataset & classes & spectral bands ($N_{c}$) & Block 1 & $N_{b}$ & patch size \\ \hline

         Indian Pines  & 11 & 220  & 1 $\times$ 1 &  4 &  3 $\times$ 3  \\ \hline

         Salinas & 16 & 224 & 1 $\times$ 1 & 8 &  3 $\times$ 3  \\  \hline

         KSC & 13 & 176 &  3 $\times$ 3 & 8  &  5 $\times$ 5  \\ \hline

         Botswana & 14 & 144 &  3 $\times$ 3 & 8 & 5 $\times$ 5 \\  \hline

    \end{tabular}
    \label{tab:config}
   
\end{table}

\subsection{Experiment Setup}

The proposed accelerator is developed using Verilog HDL. 
The hardware system is built on Xilinx Zynq ZC706 board which consists of a Xilinx XC7Z045 FPGA, dual ARM Cortex-A9 Processor and 1 GB DDR3 memory.  The whole system is implemented with Vivado Design Suite. The ARM processor in the Zynq device is used to initialize the accelerator, set the layer parameters and transfer the weights of each layer.  All designs run on a
single 250 MHz clock frequency. 


For comparison, the respective software implementations run on CPU and GPU are using the deep learning software framework TensorFlow \cite{tensorflow} in CentOS 7.2 operating system. 
The CPU platform is Intel Core Xeon 4110 CPU@2.10GHz with 8 cores. The GPU platform is NVIDIA GeForce 1080 (Pascal) with 2560 CUDA cores and 8GB GDDR5 256-bit memory).

\subsection{Classification Accuracy}
We first evaluate the overall accuracy (OA) of the proposed accelerator for the four benchmark datasets. Here the average per-class accuracy is omitted for each datasets due to lack of space. 
Tables \ref{tab:accuracy} shows the results of the comparison
of the proposed framework with other traditional and deep learning
based methods. From the table, we can see that our proposed network achieves nearly the same accuracy compared to BASS Net and even better overall accuracy for the Botswana dataset.
It is not surprising that our proposed framework outperforms all the
other traditional methods (k-NN and SVM) on all the evaluated four data sets in terms of accuracy. 

\begin{table}[h]
    \centering
     \caption{Classification accuracy (\%) comparison of the proposed network and other methods on the benchmark datasets.}
    \begin{tabular}{|P{2cm}|P{1cm}|P{2cm}|P{2cm}|P{2cm}|P{2cm}|}
    \hline
    OA (\%) & k-NN & SVM & BASS Net & Proposed \\ \hline

        Indian Pines  & 76.4 & 89.8 & 96.7 & 95.8 \\  \hline

         Salinas& 86.3 & 93.1 & 98.9& 98.9 \\  \hline

        KSC & 79.5 &  89.1 & 95.3 & 95.2 \\  \hline

        Botswana  & 81.6 & 85.4 & 98.1 &  \textcolor{blue}{98.7} \\  \hline

    \end{tabular}
    \label{tab:accuracy}
   
\end{table}

\subsection{Resource Utilization}

Table \ref{table:res} shows the resource utilization (LUTs, FFs, DSPs, etc.)
of our proposed accelerator when implemented in the target Zynq device. 
The implemented accelerator contains 64 CONV kernels and 256 FC kernels, i.e., $P_{C}=64$ and $P_{F}=256$. From Table 5, we can see that the computational resource, i.e., DSPs are almost fully utilized
and the allocation is balanced between the CONV and FC modules. The on-chip memories are sufficient to store the total amount of input data and output data, since the intermediate data size is relatively small for the proposed network (see Table \ref{tab:net}). This is because the CNN-based HSI method is doing pixel-wise processing and we can process each pixel in on-chip buffers.

\begin{table}[ht]

\centering
\caption{FPGA resource utilization of the accelerator.}

\begin{tabular}{| p{2cm} | P{1.5cm} | P{1.5cm} | P{1.5cm}| P{1.5cm} |} 
\hline
Resources & LUTs & FFs & DSPs & BRAMs \\ 
\hline

Used & 46866 & 108991 & 832 & 210 \\
\hline

Total & 218600 & 437200 & 900 & 545 \\
\hline

Utilization & 21.4\% & 24.9\% & 92.4\% & 38.5\% \\
\hline

\end{tabular}
\vspace{-1ex}
\label{table:res}
\end{table}

\subsection{Performance Comparison vs. Other Processors and Accelerators}
We then compare the performance of our FPGA-based accelerator in FPGA platform with other platforms (CPUs and GPUs). 
The CuDNN libraries and batch processing are used for optimizing the GPU solution,
and the compilation flag -Ofast is activated for the CPU implementation.
The results are shown in Table \ref{tab:speed1}. As a reference, we also show the execution time of BASS Net in CPU and GPU. The BASS Net is not implemented in FPGA platforms due to the reasons mentioned in Section 2.2. Our accelerator achieves the processing speed of 25.2, 26, 16.4 and 11.2 us/pixel respectively for the four datasets. Compared to the GPU, the average speedup is about 3 times. Compared to the CPU, we achieve more than 70x speedup. 
It should be noted that CPUs and GPUs are not realistic to be mount on a satellite
or a drone because of their high power consumption, and therefore their usability is very limited in space platforms.

\begin{table}[h]

    \centering
     \caption{Speedup of our proposed accelerators vs. CPUs and GPUs.}
     \begin{threeparttable}
    \begin{tabular}{|P{2cm}|P{1.7cm}|P{2cm}|P{2cm}|P{2cm}|}
    \hline
    \multicolumn{2}{ |c| } {Execution time (us/pixel)} & CPU & GPU & FPGA \\ \hline
         
         \rowcolor{LightCyan}
         \multirow{3}{*}{Indian Pines}  & BASS & 1166 & 123 & -   \\  \cline{2-5}
                                       & Ours  & 2180  & 99.6  &  25.2  \\ \cline{2-5}
                                         \rowcolor{Gray} 
                                       & Speedup\tnote{*} & \textcolor{blue}{86.5x} & \textcolor{blue}{3.9x} & 1x \\ \hline
           \rowcolor{LightCyan}                                      
         \multirow{3}{*}{Salinas}  & BASS & 1170 & 100.6 & -  \\  \cline{2-5}
                                & Ours  & 2026 & 102 & 26  \\ \cline{2-5}
                              \rowcolor{Gray}     & Speedup\tnote{*} & \textcolor{blue}{78x} & \textcolor{blue}{3.9x} & 1x   \\  \hline 
            \rowcolor{LightCyan}                                    
         \multirow{3}{*}{KSC}  & BASS &  723 & 49.6 & - \\  \cline{2-5}
                                       & Ours & 1511 & 46.4 & 16.4 \\ \cline{2-5}
                                    \rowcolor{Gray}    & Speedup\tnote{*} & \textcolor{blue}{92x} & \textcolor{blue}{2.8x} & 1x   \\ \hline
                                       
            \rowcolor{LightCyan}                                     
         \multirow{3}{*}{Botswana}  & BASS & 808 & 53.5 & - \\  \cline{2-5}
         
                                       & Ours  & 978  & 37.7 & 11.2 \\ \cline{2-5}
                                \rowcolor{Gray}     & Speedup\tnote{*} & \textcolor{blue}{87x} & \textcolor{blue}{3.4x} & 1x  \\ \hline

    \end{tabular}
    \begin{tablenotes}
  
       \item[*] \small{The numbers in this row represent the speedups of our model run in FPGA platform compared to that run in CPU and GPU platforms.}

    \end{tablenotes}
    \end{threeparttable}
    \label{tab:speed1}

\end{table}

Finally we compare our accelerator to other FPGA-based accelerators implementing SVM \cite{arc_ws,fpt2018}. 
These two accelerator are implemented in an Altera Stratix
V 5SGSMD8N2F45C2 FPGA on Maxeler MAX4 DFE \cite{arc_ws}, and in a Xilinx Kintex-7
XC7K325T-FF2-900 FPGA device \cite{fpt2018} respectively. Due to these accelerators have different DSP numbers compared to ours, we also provide the speedups normalized by the number of DSPs. The results are shown in Table \ref{tab:speed2}.

\begin{table}[h]
    \centering
     \caption{Accuracy, Speed and Power consumption of our accelerator vs. other FPGA accelerators implemented based on SVM methods.}
     \begin{threeparttable}
    \begin{tabular}{|P{2.7cm}|P{2.2cm}|P{1.5cm}|P{2.2cm}|P{1.2cm}|P{1.2cm}|}
    \hline
    
     \multirow{2}{*}{} & \multirow{2}{*}{ Accuracy (\%) } & \multicolumn{3}{ c| }{Speed } & Power \\ \cline{3-5}
                       &  & Mpixels/s & Kpixels/s/DSP & \#DSP & (W) \\ \hline
     
    SVM DFE \cite{arc_ws} & 85.4  & 1.01 & 0.6 & 1680 & 26.3  \\ \hline

     SVM Kintex-7 \cite{fpt2018}\  & 81.3  & 0.65 & 0.7  & 840  &  4.25 \tnote{a}  \\ \hline
     
     \rowcolor{Gray}   Ours & 98.7 & 0.09 & 0.1 & 900  & 9.0 \tnote{b}  \\ \hline

    \end{tabular}
         \begin{tablenotes}
  
       \item[a] \small{This is the power consumption reported in \cite{fpt2018}.}
\item[b] \small{The power consumption of ours is measured from the board using a power meter.}

    \end{tablenotes}
    \end{threeparttable}
    \label{tab:speed2}
\vspace{-1ex}
\end{table}

Compared to the SVM-based accelerators, our accelerator is based on CNN model and therefore is much more computationally intensive due to the large complexity of the network in order for accuracy improvement. This is exactly the motivation that we accelerate the CNN-based HSI models on FPGA platforms. Nevertheless, our accelerator still achieves the same scale of speed in terms of both pixel per second and pixel per second per DSP. Besides, our accelerator has much less power consumption than the accelerator in \cite{arc_ws}. Therefore, our proposed accelerator is more promising for embedded HSI applications with high accuracy and low power consumption requirement and on-board platforms.



%% file: Summary.tex
\section{Conclusion}

This work proposes a hardware accelerator for CNN-based HSI applications on FPGA platforms. We first adapt the state-of-the-art BASS Net for hardware efficiency and accuracy improvement. Then we propose a hardware architecture to accelerate our proposed network to achieve real-time processing speed. Hardware optimization techniques are applied to customize and optimize our accelerator together with design space exploration.
Experimental results based on public datasets show that our accelerator achieves notable accuracy improvement compared to previous SVM-based FPGA accelerators, and significant speedup compared to the respective implementation of our CNN model on CPUs and GPUs. 
Future work includes extending the network with other types of layers such as depth-wise convolution for enhancing accuracy and performance, and developing automatic tools to generate hardware designs for HSI applications.

\section*{Acknowledgement}
The support of the UK EPSRC (EP/I012036/1, EP/L00058X/1, EP/L016796/1 and EP/N031768/1), the European Union Horizon 2020 Research and Innovation Programme under grant agreement number 671653, Altera, Corerain, Intel, Maxeler, SGIIT, and the China Scholarship Council is gratefully acknowledged. 

%% file: main.bbl
\begin{thebibliography}{10}
\providecommand{\url}[1]{\texttt{#1}}
\providecommand{\urlprefix}{URL }
\providecommand{\doi}[1]{https://doi.org/#1}

\bibitem{tensorflow}
Abadi, M., et~al.: {TensorFlow: Large-Scale Machine Learning on Heterogeneous
  Systems} (2015), \url{https://www.tensorflow.org/}

\bibitem{bioucas2013hyperspectral}
Bioucas-Dias, J.M., et~al.: Hyperspectral remote sensing data analysis and
  future challenges. IEEE Geoscience and remote sensing magazine
  \textbf{1}(2),  6--36 (2013)

\bibitem{chen2014deep}
Chen, Y., Lin, Z., Zhao, X., Wang, G., Gu, Y.: Deep learning-based
  classification of hyperspectral data. IEEE Journal of Selected topics in
  applied earth observations and remote sensing  \textbf{7}(6),  2094--2107
  (2014)

\bibitem{grahn2007techniques}
Grahn, H., Geladi, P.: Techniques and applications of hyperspectral image
  analysis. John Wiley \& Sons (2007)

\bibitem{lee2017going}
Lee, H., Kwon, H.: Going deeper with contextual cnn for hyperspectral image
  classification. IEEE Transactions on Image Processing  \textbf{26}(10),
  4843--4855 (2017)

\bibitem{Leng2016CubeCNNSVMAN}
Leng, J., et~al.: Cube-cnn-svm: A novel hyperspectral image classification
  method. 2016 IEEE 28th International Conference on Tools with Artificial
  Intelligence (ICTAI) pp. 1027--1034 (2016)

\bibitem{liu2017communication}
Liu, S., Bouganis, C.S.: {Communication-Aware MCMC Method for Big Data
  Applications on FPGAs}. In: IEEE International Symposium on
  Field-Programmable Custom Computing Machines (FCCM). pp. 9--16 (2017)

\bibitem{liu2014parallel}
Liu, S., Mingas, G., Bouganis, C.S.: {Parallel resampling for particle filters
  on FPGAs}. In: IEEE International Conference on Field-Programmable Technology
  (FPT). pp. 191--198 (2014)

\bibitem{liu2015exact}
Liu, S., Mingas, G., Bouganis, C.S.: {An exact MCMC accelerator under custom
  precision regimes}. In: IEEE International Conference on Field Programmable
  Technology (FPT). pp. 120--127 (2015)

\bibitem{liu2017unbiased}
Liu, S., Mingas, G., Bouganis, C.S.: {An unbiased mcmc fpga-based accelerator
  in the land of custom precision arithmetic}. IEEE Transactions on Computers
  \textbf{66}(5),  745--758 (2017)

\bibitem{liu2017trets}
Liu, S., et~al.: {Optimizing CNN-based Segmentation with Deeply Customized
  Convolutional and Deconvolutional Architectures on FPGA}. ACM Transactions on
  Reconfigurable Technology and Systems (TRETS)  (2018)

\bibitem{lopez2013promise}
Lopez, S., et~al.: The promise of reconfigurable computing for hyperspectral
  imaging onboard systems: A review and trends. Proceedings of the IEEE
  \textbf{101}(3),  698--722 (2013)

\bibitem{luo2018hsi}
Luo, Y., et~al.: Hsi-cnn: A novel convolution neural network for hyperspectral
  image. In: 2018 International Conference on Audio, Language and Image
  Processing (ICALIP). pp. 464--469. IEEE (2018)

\bibitem{HSIEnviroMonitor}
Martin, M.E., , et~al.: Development of an advanced hyperspectral imaging (hsi)
  system with applications for cancer detection. Annals of biomedical
  engineering  \textbf{34}(6),  1061--1068 (2006)

\bibitem{HSICancer}
Martin, M.E., Wabuyele, M.B., et~al.: Development of an advanced hyperspectral
  imaging (hsi) system with applications for cancer detection. Annals of
  Biomedical Engineering  \textbf{34}(6),  1061--1068 (2006),
  \url{https://doi.org/10.1007/s10439-006-9121-9}

\bibitem{salem2001hyperspectral}
Salem, F., et~al.: Hyperspectral image analysis for oil spill detection. In:
  Summaries of NASA/JPL Airborne Earth Science Workshop, Pasadena, CA. pp.~5--9
  (2001)

\bibitem{santara2017bass}
Santara, A., et~al.: Bass net: band-adaptive spectral-spatial feature learning
  neural network for hyperspectral image classification. IEEE Transactions on
  Geoscience and Remote Sensing  \textbf{55}(9),  5293--5301 (2017)

\bibitem{fpt2018}
Tajiri, K., Maruyama, T.: Fpga acceleration of a supervised learning method for
  hyperspectral image classification. In: 2018 International Conference on
  Field-Programmable Technology (FPT). IEEE (2018)

\bibitem{arc_ws}
Wang, S., Niu, X., Ma, N., Luk, W., Leong, P., Peng, Y.: A scalable dataflow
  accelerator for real time onboard hyperspectral image classification. In:
  Applied Reconfigurable Computing. pp. 105--116. Springer International
  Publishing, Cham (2016)

\bibitem{zhang2016deep}
Zhang, L., Zhang, L., Du, B.: Deep learning for remote sensing data: A
  technical tutorial on the state of the art. IEEE Geoscience and Remote
  Sensing Magazine  \textbf{4}(2),  22--40 (2016)

\bibitem{zhao2017optimizing}
Zhao, R., Niu, X., Wu, Y., Luk, W., Liu, Q.: Optimizing cnn-based object
  detection algorithms on embedded fpga platforms. In: ARC. pp. 255--267.
  Springer (2017)

\end{thebibliography}
